\documentstyle[12pt]{article}

\def\ftoday{{\sl  \number\day \space\ifcase\month 
\or Janvier\or F\'evrier\or Mars\or avril\or Mai
\or Juin\or Juillet\or Ao\^ut\or Septembre\or Octobre
\or Novembre \or D\'ecembre\fi
\space  \number\year}}    
\newcommand{\journal}[4]{{\em #1~}#2\,(19#3)\,#4;}

\newcommand{\hpa}{\journal {Helv. Phys. Acta}}

\newcommand{\ijmp}{\journal {Int. J. Mod. Phys.}}

\newcommand{\pr}{\journal {Phys. Rev.}}

\newcommand{\jmp}{\journal {J. Math. Phys.}}

\newcommand{\cmp}{\journal {Commun. Math. Phys.}}

\newcommand{\cqg}{\journal {Class. Quantum Grav.}}

\newcommand{\np}{\journal {Nucl. Phys.}}
\newcommand{\pl}{\journal {Phys. Lett.}}

\newcommand{\prep}{\journal {Phys. Rep.}}

\makeatletter
\@addtoreset{equation}{section}
\makeatother

\newcommand{\es}{\\[3mm]}

\renewcommand{\a}{\alpha}

\newcommand{\g}{\gamma}           \newcommand{\G}{\Gamma}
\renewcommand{\d}{\delta}

\newcommand{\m}{\mu}
\newcommand{\n}{\nu}

\newcommand{\th}{\theta}         
\newcommand{\f}{{\phi}}           

\newcommand{\BB}{{\cal B}}

\newcommand{\NN}{{\cal N}}

\newcommand{\SS}{{\cal S}}

\newcommand{\ch}{\widehat{C}}
\newcommand{\gh}{\widehat{\gamma}}
\newcommand{\W}{W_{i}}

\newcommand{\sla}{\raise.15ex\hbox{$/$}\kern -.57em} 
\newcommand{\Sla}{\raise.15ex\hbox{$/$}\kern -.70em}

\def\Lp{\displaystyle{\biggl(}}
\def\Rp{\displaystyle{\biggr)}}
\def\LP{\displaystyle{\Biggl(}}
\def\RP{\displaystyle{\Biggr)}}

\newcommand{\complex}{{\kern .1em {\raise .47ex
\hbox {$\scriptscriptstyle |$}}
    \kern -.4em {\rm C}}}
\newcommand{\real}{{{\rm I} \kern -.19em {\rm R}}}
\newcommand{\rational}{{\kern .1em {\raise .47ex
\hbox{$\scripscriptstyle |$}}
    \kern -.35em {\rm Q}}}
\renewcommand{\natural}{{\vrule height 1.6ex width
.05em depth 0ex \kern -.35em {\rm N}}}
\newcommand{\tint}{\int d^2 \! x \,  d^2 \! \th \, }
\newcommand{\intg}{\int d^D \! x \, }

\newcommand{\half}{\dfrac{1}{2}}

\newcommand{\pad}[2]{{\frac{\partial #1}{\partial #2}}}
\newcommand{\fud}[2]{{\frac{\delta #1}{\delta #2}}}

\newcommand{\dfrac}[2]{{\displaystyle{\frac{#1}{#2}}}}

\newcommand{\twiddle}{\lower.9ex\rlap{$\kern -.1em\scriptstyle\sim$}}

\newcommand{\equ}[1]{(\ref{#1})}
\newcommand{\eq}{\begin{equation}}
\newcommand{\eqn}[1]{\label{#1}\end{equation}}
\newcommand{\eea}{\end{eqnarray}}
\newcommand{\eqa}{\begin{eqnarray}}
\newcommand{\eqan}[1]{\label{#1}\end{eqnarray}}
\newcommand{\ba}{\begin{array}}
\newcommand{\ea}{\end{array}}
\newcommand{\eqac}{\begin{equation}\begin{array}{rcl}}
\newcommand{\eqacn}[1]{\end{array}\label{#1}\end{equation}}

\begin{document}
\newcommand{\cb}{{\bar c}}
\newcommand{\mn}{{\m\n}}
\newcommand{\pic}{$\spadesuit\spadesuit$}
\newcommand{\?}{{\bf ???}}

\titlepage  \noindent
{
GEF-TH-12/1998 \vspace{8mm}

\begin{center} 
{\huge  Repairing Broken Algebras }

\vspace{2cm}

{\Large Alberto Blasi and Nicola Maggiore}

{\it Dipartimento di Fisica -- Universit\`a di Genova\\
via Dodecaneso 33 -- I-16146 Genova\\Italy}

\vspace{15mm}

\end{center}

\begin{center}
{\bf Abstract}
\end{center}
{\it We consider theories characterized by a set of Ward operators 
which do not form a closed algebra. We impose the Slavnov--Taylor 
identity built out of the Ward operators and we derive 
the acceptable breaking of the algebra and 
the general 
form of the classical 
action. The 1PI generating functional is expressed in terms of the 
known quantities characterizing the theory and of a nontrivial integrability 
condition. As a nontrivial application of our formalism, we discuss 
the N=4 supersymmetric nonlinear sigma model.}

\vfill\noindent
}
\newpage
\section{Introduction}
The approach to symmetries $via$ the introduction of anticommuting 
ghosts and hence the construction of a BRS nilpotent operator has 
provided a powerful tool to analyze the renormalization of quantum field 
theories with nonlinearly 
realized symmetries, both local and global~\cite{brs} 

It was soon realized that the extension of the original symmetric 
action to one containing external fields  coupled to the BRS 
variations of the quantum fields, made possible to write the BRS 
operator in a kind of universal way as a Slavnov--Taylor identity
making in addition possible the  control of the 
behavior under renormalization of the symmetry itself. Furthermore, 
as  extra bonus, the bilinear terms in the sources, 
if the power counting and ghost number allow these 
insertions in the action,  account for the presence of symmetries 
realized only on--shell, $i.e.$ modulo the equations of motion of 
the fields~\cite{bv}. 

Within this framework, the renormalization program 
reduces, thanks to the Quantum Action Principle~\cite{qap}, 
to an algebraic 
discussion of the cohomology spaces of the nilpotent BRS operator. 

Our aim in this paper is that of discussing and putting into evidence 
the information we can obtain from a nilpotent operator: we shall 
therefore define a BRS operator which is only related to the 
covariance--and not necessarily to the invariance--properties of a 
classical action under a set of nonlinear transformations of the 
quantum fields. These nonlinear transformations are not required to 
be a realization of some Lie algebra either, but nonetheless satisfy 
algebraic relations, obtained through their commutators and/or 
anticommutators. A step in this directions was already taken in 
\cite{beppe}, where it was described the mechanism of reconstructing 
a Lie algebra from a nilpotent operator and the consequences for 
the cohomology spaces. 

Since our goal is to discuss the most general 
case, we try to keep the computational complexities to a 
minimum, and shall therefore work only with scalar fields 
and global ghosts, just to illustrate our method,  having in mind the 
action of supersymmetric non linear sigma model in two spacetime 
dimensions, which is explicitly 
discussed as an example.
In the same 
spirit, we shall not put any power counting constraint to the external 
field sector of the classical action, and thus we can investigate in 
full generality the meaning of their contribution.

The paper is organized as follows. 

In Section 2, we shall 
obtain some algebraic relations, which will leads us to the most 
general form of the broken algebra and of the classical action 
compatible with the Slavnov--Taylor identity. Moreover, we shall find 
a condition, derived from the external field sector of the action, whose 
interpretation is not straightforward, but which is verified in all 
the examples known in the Literature. Particularly relevant is the 
case of $D=2$, $N=4$ 
supersymmetric nonlinear sigma model, which  is analyzed in Section 3.

In Section 4, we shall study the conditions under which it is possible 
to recover a Lie algebra structure for the theory, and we shall find 
out several possibilities for that. We shall also see that 
the nilpotent operator one can construct in the generic case of 
transformations not related to a Lie algebra invariance of the 
classical action, leads to a trivial cohomological problem. When we 
restrict to a Lie algebra invariance, we shall recover a possible 
non-triviality of the cohomology spaces and we shall also find that the 
on--shell realization of the symmetry is indeed the most general 
case we can have. 

Our conclusions are finally drawn in Section 5.
\section{Imposing the Slavnov--Taylor Identity 
}\label{sez2}
Consider a classical action $I(\f^{a})$, built on a flat spacetime of 
generic dimension $D$ and depending on a number of fields 
$\{\f^{a}(x)\}$ --which for simplicity we assume to be scalar-- 
labeled by an index $a$. Let $\{W_{i}\}$ be a set of Ward operators 
described by the functionals
\eq
W_{i} =\intg W^{a}_{i}(\f)\fud{}{\f^{a}} 
\eqn{W}
where $W^{a}_{i}(\f)$ depends on the fields in general in a nonlinear 
way. The Ward operators~\equ{W} are the functional realizations of the 
nonlinear field transformations
\eq
\d_{i}\f^{a}(x) = W^{a}_{i}(\f)
\eqn{transf}
We are interested in treating the case in which the operators $W_{i}$ do 
not form a closed algebra. The most general case is indeed given by 
the following structure
\eq
[W_{i},W_{j}]=D_{ij}
\eqn{algebra}
The expression~\equ{algebra} covers all possible algebraic 
structures, including the closed ones, obtained when the 
antisymmetric operator $D_{ij}$ is given by
\eq
D_{ij} = f_{ijk} W_{k}
\eqn{closed}
for some structure constants $f_{ijk}$.

What is more difficult to treat, is the case of open algebras. In this 
case indeed $D_{ij}$ represents an obstruction. Examples of such open 
structures are given by the supersymmetry algebra, where 
the commutator between two supersymmetries finds as obstructions to 
the closure on traslations, the equations of motion and the gauge 
transformations~\cite{sw}. Another relevant example of open algebras, are the 
reducible symmetries of Batalin -- Vilkovisky~\cite{bv}, 
characterizing for 
instance a class of topological models~\cite{top}. In such theories, one 
lands on a BRS operator which is nilpotent only once the equations of 
motion are used.

We would like here to work on a very general ground, without 
referring to a particular model. we shall find the most general form 
of the breaking $D_{ij}$ in ~\equ{algebra}, embedded in the 
Slavnov--Taylor identity holding for the 1PI generating functional~$\G$.

A convenient way to proceed is to introduce global ghosts $C^{i}$ 
and $\lambda$ in order to write a nilpotent BRS operator
\eq\ba{l}
s = C^{i}W_{i} + \lambda\half C^{i}C^{j}D_{ij} - \pad{}{\lambda} \es
\phantom{s} = W + \lambda d - \pad{}{\lambda}
\ea\eqn{brs}
where we defined 
\eq
W\equiv C^{i}W_{i}\ \ \ d\equiv \half C^{i}C^{j}D_{ij}
\eqn{d}
 The nilpotency of the BRS operator $s$ is insured 
by the identities $W^{2}=0$ and $[W,d]=0$ and by the fact that the 
ghosts $C^{i}$ and $\lambda$ are anticommuting grassmannian variables, to 
which we assign charge $+1$ and $-1$ respectively, in order to have a 
BRS operator raising by one unit the ghost charge. The 
Slavnov--Taylor (ST) operator corresponding to the BRS 
operator~\equ{brs} 
reads
\eq
\SS(\G^{cl}) = \intg \fud{\G^{cl}}{\g_{a}(x)} \fud{\G^{cl}}{\f^{a}(x)}
+ \pad{\G^{cl}}{\lambda} 
\eqn{slavnov}
In \equ{slavnov}, $\G^{cl}$ is the tree level 1PI generating 
functional, whose most general form is
\eq\ba{l}
\G^{cl} = I(\f^{a}) + \intg \LP
\g_{a}C^{i}W^{a}_{i}(\f) + 
\g_{a}\g_{b}C^{i}C^{j}W^{ab}_{ij}(\f) +
\lambda C^{i}A_{i}(\f)  \es
\phantom{\G^{cl} = I(\f^{a}) + \intg \LP}
+ \lambda\g_{a}C^{i}C^{j}X^{a}_{ij}(\f) +
\lambda\g_{a}\g_{b}C^{i}C^{j}C^{k}X^{ab}_{ijk}(\f) \RP 
\ea\eqn{gamma}
A few comments on the terms appearing in $\G^{cl}$ are in order~:
\begin{enumerate}
\item
The $\g_{a}(x)$ are external sources coupled to the nonlinear 
variations $W^{a}_{i}(\f)$ of the quantum fields $\f^{a}(x)$, according 
to the standard method of treating nonlinear symmetries;
\item
For closed algebras, only the term linear in the external sources 
appears in $\G^{cl}$. $W_{ij}^{ab}(\f)$, $A_{i}(\f)$, $X^{a}_{ij}(\f)$, 
and $X^{ab}_{ijk}(\f)$ are generic polynomials in the fields 
$\f^{a}(x)$, constrained only by the fulfillment of the ST identity
\eq
\SS(\G^{cl}) = 0
\eqn{slavnovid}
\item
We demand the conservation of the ghost charge. In order to have 
$\G^{cl}$ uncharged, the sources $\g_{a}(x)$ must be assigned 
charge $-1$;
\item
Since we are considering arbitrary spacetime dimensions and we are not 
referring to a particular model, we do not impose any power counting 
restriction on $\G^{cl}$. Nonetheless, we restrict ourselves to terms 
at most quadratic in the external sources, to keep contact with the 
cases of most concern. Our analysis can be trivially extended to the 
case of higher powers in the $\g_{a}$'s.
\end{enumerate}
Introducing the notation
\eq\ba{l}
C^{i}C^{j}W^{ab}_{ij}(\f) \equiv W^{ab}(C,\f) \es
C^{i}A_{i}(\f)  \equiv A(C,\f) \es
C^{i}C^{j}X^{a}_{ij}(\f) \equiv X^{a}(C,\f) \es
C^{i}C^{j}C^{k}X^{ab}_{ijk}(\f) \equiv  X^{ab}(C,\f)
\ea\eqn{def}
the action $\G^{cl}$ can be written in a more compact way~:
\eq 
\G^{cl} = I(\f^{a}) + \intg \LP
\g_{a}W^{a} + 
\g_{a}\g_{b}W^{ab} +
\lambda C^{i}A_{i} 
+ \lambda\g_{a}X^{a} +
\lambda\g_{a}\g_{b}X^{ab} \RP
\eqn{action} 
It is useful to summarize in a table the ghost charges ($\Phi\Pi$) 
involved~: 
\begin{table}[hbt]
\centering
\begin{tabular}{|c||c|c|c|c|c|c|c|c|c|}
\hline
&$\f^{a}$&$\g_{a}$&$\lambda$&$W^{a}$&$W^{ab}$&$A$&$X^{a}$&$X^{ab}$&$d$ \\
\hline\hline
$\Phi\Pi$&0&-1&-1&1&2&1&2&3&2 \\
\hline
\end{tabular}
\label{table}
\end{table}

Our approach is to require that the action $\G^{cl}$ \equ{action} 
satisfies the ST identity \equ{slavnovid}. In this way we shall find 
the most general form of the breaking $d$~\equ{d} and of the functions 
$W^{ab}(C,\f)$,  $A(C,\f)$, $X^{a}(C,\f)$ and $X^{ab}(C,\f)$, which 
for the moment are left generic.
Recall that, the only known elements are the variations 
of the fields $W_{a}(\f)$ and the classical action $I(\f)$.

The calculation, although rather lengthy, is straightforward. It is 
convenient to analyze the ST identity according to the powers of 
$\g_{a}(x)$ and $\lambda$. 

At the zero order, we find that the ST identity \equ{slavnovid} is 
satisfied provided that
\eq
\intg \LP W^{a}\fud{I}{\f^{a}}+ A \RP = 0
\eqn{aexpr}
as it to say that the functional $A_{i}(\f)$ in \equ{gamma} can be 
interpreted as a breaking term. In other words, in our formalism we do 
not ask that the Ward operators $W_{i}$ describe symmetries of the 
classical action $I(\f)$, but we allow for a (in general nonlinear) 
breaking $A_{i}(\f)$.

Considering the order $O(\g)$, we find the expression for 
$X^{a}(C,\f)$~:
\eq
X^{a}(C,\f) = 2 W^{ab}(C,\f)\fud{I(\f)}{\f^{b}} + 
W^{b}(C,\f)\fud{W^{a}(C,\f)}{\f^{b}}
\eqn{xaexpr}
in terms of the known $W^{a}(C,\f)$ and the unknown $W^{ab}(C,\f)$. 
Applying to both sides of \equ{xaexpr} $\fud{}{\f^{a}}$, summing over 
$a$, performing the spacetime integration and remembering that 
$W^{2}=d$, we find
\eq
d = \intg \LP X^{a} - 2 W^{ab} \fud{I}{\f^{b}} \RP\fud{}{\f^{a}}
\eqn{dexpr}
The equation \equ{dexpr} represents the most general form of 
$d$~\equ{d}, as 
results from the imposition of the ST identity \equ{slavnovid}, 
written in terms of quantities appearing in the classical action 
$\G^{cl}$. The breaking of the algebra \equ{algebra}
consists therefore of two parts, one of which is vanishing once the 
equations of motion are satisfied ($\fud{I}{\f^{a}}=0$). The other 
term is $\intg X^{a}\fud{}{\f^{a}}$, and is still there even 
on--shell, representing the bulk of the breaking. The supersymmetry 
algebra~\cite{sw} is an example of an algebra whose breaking has 
exactly the structure described in \equ{dexpr}. The form of $d$ 
provides moreover a simple interpretation of the terms $X^{a}(C,\f)$ and 
$W^{ab}(C,\f)$ appearing in \equ{action}. We recall that, while the 
functional $X^{a}(C,\f)$ is determined by \equ{xaexpr}, $W^{ab}(C,\f)$ 
is still completely unconstrained.

We go on imposing the ST identity, by selecting the term quadratic in 
the external fields, and we find the following expression for 
$X^{ab}(C,\f)$~:
\eq
X^{ab}(C,\f) = 
W^{ac}(C,\f) \fud{W^{b}(C,\f)}{\f^{c}}
- W^{bc}(C,\f) \fud{W^{a}(C,\f)}{\f^{c}}
- W^{c}(C,\f) \fud{W^{ab}(C,\f)}{\f^{c}}
\eqn{xabexpr}

Finally, the ST identity at the order $O(\g^{3})$ is satisfied 
provided that
\eq
W^{ab}(C,\f) \fud {W^{cd}(C,\f)}{\f^{a}} 
+ W^{ad}(C,\f) \fud {W^{bc}(C,\f)}{\f^{a}}
+ W^{ac}(C,\f) \fud {W^{db}(C,\f)}{\f^{a}}
= 0
\eqn{wexpr}
All the other constraints deriving from the ST identity are 
automatically satisfied once the expressions found for $A(C,\f)$ 
\equ{aexpr}, $X^{a}(C,\f)$ \equ{xaexpr}, $X^{ab}(C,\f)$ \equ{xabexpr}, the 
constraint on $W^{ab}(C,\f)$ \equ{wexpr} and the algebraic relation 
$[W,d]=0$, hold.

At this point, every term in the action $\G^{cl}$ \equ{gamma} is 
completely determined in terms of $I(\f)$, $W^{a}_{i}(\f)$ and 
$W^{ab}_{ij}(\f)$. Now, while the first two quantities are given, 
being our starting point, the functional $W^{ab}_{ij}(\f)$ is not 
determined by the condition \equ{wexpr}. Looking at the general 
expression we found for the operator $d$ \equ{dexpr}, we realize that 
the $W^{ab}$'s are related to that part of the breaking of the 
algebra \equ{algebra} which is vanishing on--shell. In most explicit 
cases, the $W^{ab}$'s do not depend on the fields, and the 
constraint \equ{wexpr} is trivially satisfied. For a nontrivial field 
dependence of $W^{ab}$, the relation \equ{wexpr} can be interpreted 
as an integrability condition, whose validity is not straightforwardly 
guaranteed. In the next Section, in fact, we discuss a case in which 
\equ{wexpr} has an evident geometrical meaning, not directly related 
to the action and its symmetries.

Once the ST identity \equ{slavnovid} is satisfied, one can verify 
that the nilpotency of the corresponding linearized ST operator
\eq
\BB_{\G^{cl}} = \intg 
\LP 
\fud{\G^{cl}}{\f^{a}(x)}\fud{}{\g_{a}(x)}
+
\fud{\G^{cl}}{\g_{a}(x)}\fud{}{\f^{a}(x)}
\RP
+ \pad{}{\lambda}
\eqn{slavnovlin}
is guaranteed.

\section{An example: $D=2$, $N=4$ supersymmetric nonlinear sigma model}
\label{sez3}
In most cases the coefficients $W^{ab}(C,\phi)$ are field 
independent, and hence trivially satisfy \equ{wexpr}; one remarkable 
exception is the $D=2$, $N=4$ supersymmetric nonlinear sigma 
model, described, in absence of torsion, by the action
\eq
I(\phi^{a}) = \tint 
g_{ab}(\phi)[D_{+}\phi^{a}D_{-}\phi^{b}]
\eqn{sigma-action}
The invariant action \equ{sigma-action} is written in terms of 
light-cone coordinates, and
\eq 
D_{\pm}=\pad{}{\th^{\pm}} + i\, \th^{\pm}\pad{}{x^{\pm}}
\eqn{derivative} 
Moreover, depending on $N=1$ superfields $\phi^{a}(x,\th)$, the 
action is manifestly \mbox{$N=1$} 
supersymmetric. Nevertheless it has been shown~\cite{agf}, that
the action \equ{sigma-action} is invariant under the following additional three 
nonlinear (super)field transformations
\eq\ba{l}
\delta \phi^{a} = J^{a}_{ib}(\phi) 
[\epsilon^{+}_{i} D_{+}\phi^{b} + 
  \epsilon^{-}_{i} D_{-}\phi^{b}] \es
\phantom{\delta \phi^{a}} \equiv [\epsilon^{+}_{i} W_{i+} + 
                                      \epsilon^{-}_{i} 
                                      W_{i-}]\phi^{a}\ \ \ i=1,2,3
\ea\eqn{susy-trans}
provided that the tensors $J^{a}_{ib}(\phi)$ are  complex structures 
satisfying the SU(2) quaternionic relations
\eq
J^{a}_{ic}(\phi) J^{c}_{jb}(\phi) = - \delta_{ij} \delta^{a}_{b}
+ \epsilon_{ijk} J^{a}_{kb}(\phi) 
\eqn{quaternions}
and moreover obey
\eq
J^{ab}_{i}(\phi)  = J^{a}_{ic}(\phi)  g^{cb}(\phi)  = - J^{ba}_{i}(\phi) 
\eqn{hermiticy}
\eq
D_{c}J^{a}_{ib} (\phi) \equiv \partial_{c}J^{a}_{ib}(\phi)  + 
\Gamma^{a}_{cd}[g] 
J^{d}_{ib}(\phi)  - \G^{d}_{cb}[g] J^{a}_{id}(\phi)  = 0
\eqn{covariantconst}
Equation \equ{quaternions}, together with \equ{hermiticy}, implies
 that the metric is hermitian with 
respect to the $J$'s, and the identity \equ{covariantconst} 
represents the fact that the complex structures $J$ are covariantly 
constant with respect to the metric ($\G^{a}_{bc}[g]$ is the 
Christoffel connection).

The conditions \equ{quaternions}, \equ{hermiticy}  are local properties  
which can be globally extended if and only if the complex structures 
$J$ are such that the corresponding Nijenhuis tensors vanish
\eq
N^{c}_{iab} = 
J^{d}_{ia}  ( \partial_{d} J^{c}_{ib} - \partial_{b} 
J^{c}_{id} ) - ( a\leftrightarrow b ) = 0 
\eqn{nij}
Now, the condition \equ{covariantconst} implies the vanishing of the 
Nijenhuis tensor \equ{nij}, so that the properties \equ{quaternions} 
and \equ{hermiticy} can be extended to the whole Riemannian 
manifold spanned by the  
coordinates $\phi^{a}$. This defines the 
manifold to be of hyperk\"ahler type~\cite{math}. In other words, the 
existence of extended supersymmetries for the action 
\equ{sigma-action} requires that the target space is an 
hyperk\"ahler Riemannian manifold~\cite{agf}. The quantization of 
this model has been performed in \cite{bon}.

The supersymmetry algebra closes only on--shell, since we have
\eq
\{W_{i+}, W_{j-}\}\phi^{a}= \epsilon_{ijk}J^{ab}_{k} 
\fud{I(\phi)}{\phi^{b}}
\eqn{susy-algebra}
which is of the general type \equ{algebra}.

According to our analysis, we have to introduce in the total action 
terms linear in the external fields $\g_{a}$,  and also terms 
bilinear in the external fields, which, in the notation of \equ{action} 
and recalling the most general expression~ \equ{dexpr} found for the 
breaking of the algebra, 
are respectively given by
\eq
W^{a} = J^{a}_{ib}(\phi) 
[C^{+}_{i} D_{+}\phi^{b} + 
  C^{-}_{i} D_{-}\phi^{b}]
\eqn{susywa}
\eq
W^{ab} = -\half \epsilon_{ijk} J^{ab}_{k} C^{+}_{i} C^{-}_{j}
\eqn{susywab}
where $C^{\pm}_{i}$ are the commuting supersymmetric Faddeev--Popov 
constant ghosts.

The analysis of relation \equ{wexpr} with $W^{ab}(\phi)$ specified as 
above requires a good amount of algebraic patience but in the end one 
finds that it reduces to
\eq
(\a \d_{ij}  
+  \a_{ij} ) J^{da}_{i} \partial_{d} J^{bc}_{j} = 0
\eqn{result.1}
where $\a$ and $\a_{ij}$ are constant tensors
\eq\ba{l}
\a = (C^{+}_{i}C^{-}_{j} - C^{+}_{j}C^{-}_{i}) 
     (C^{+}_{i}C^{-}_{j} - C^{+}_{j}C^{-}_{i})
\es
\a_{ij} = 
2\, (C^{+}_{k}C^{-}_{i}-C^{+}_{i}C^{-}_{k})(C^{+}_{k}C^{-}_{j}-C^{+}_{j}C^{-}_{k})
\ea\eqn{result.2}
and  the left-hand side of the identity~\equ{result.1} is to be understood as 
antisymmetrized in the indices $(a,b,c)$.

Hence, the complex structures $J^{a}_{ib}(\phi)$ must 
satisfy the condition
\eq
 J^{da}_{k} \partial_{d} J^{bc}_{j} +
 J^{dc}_{k} \partial_{d} J^{ab}_{j} +
 J^{db}_{k} \partial_{d} J^{ca}_{j} = 0
\eqn{result} 
It is a remarkable fact that such integrability condition holds true  
due to the validity of 
\equ{hermiticy} and \equ{nij}. In conclusion, the identity 
\equ{wexpr}, which holds true in the generic case of open algebras, 
when applied to the particular case of $N=4$ supersymmetric nonlinear 
sigma model, shows its geometrical meaning~\equ{result}. 
It is interesting to notice that the 
vanishing of the Nijenhuis tensor, $i.e.$ the possibility of defining 
globally the complex structures, is a needed property in order to have 
a nilpotent operator, which therefore also carries information on the 
global geometrical properties of the manifold.

\section{Recovering Algebras}\label{sez4}
Once we have determined the most general action $\G^{cl}$ and the 
broken algebra compatible with the ST identity \equ{slavnovid}, the 
next natural step is to investigate whether it is possible to absorb 
the ghost $\lambda$ and to find out which are the (broken) algebraic 
structures which allow for the absorption.

The ghost $\lambda$ can be absorbed in three ways
\begin{enumerate}
\item through a redefinition of the ghosts $C^{i}$
\eq
\ch^{i} \equiv C^{i} + \lambda f^{i}_{jk}C^{j}C^{k}
\eqn{chat}
\item through a redefinition of the external fields
\eq
\gh_{a} \equiv \g_{a} + f_{ai}^{b}\lambda\g_{b}C^{i}
\eqn{ghat}
\item through a combination of \equ{chat} and \equ{ghat}
\end{enumerate}

Redefining the ghosts $C^{i}$ as in \equ{chat}, the action reads
\eq
\widehat{\G}^{cl} = I + \intg \LP 
\g_{a}\ch^{i}\widehat{W}^{a}_{i} +
\g_{a}\g_{b}\ch^{i}\ch^{j}\widehat{W}^{ab}_{ij} \RP
\eqn{gammahat}
where $\widehat{W}^{a}_{i}(\f)$ and $\widehat{W}^{ab}_{ij}(\f)$ are new 
functionals of the fields $\f^{a}$ to be determined by means of the 
functionals appearing in the previous expression for $\G^{cl}$ 
\equ{gamma}. Indeed, writing back in \equ{gammahat} the ghosts 
$C^{i}$ and identifying term by term the expressions for 
$\widehat{\G}^{cl}$ and for $\G^{cl}$, one gets
\eq\ba{l}
W^{a}_{i} = \widehat{W}^{a}_{i} \es
W_{ij}^{ab} = \widehat{W}_{ij}^{ab} \es
X_{ij}^{a} = - f^{k}_{ij}  \widehat{W}^{a}_{k} \es
A_{i} = 0 \es
X^{ab}_{ijk} = f^{l}_{jk}\widehat{W}^{ab}_{il} +
               f^{l}_{ij}\widehat{W}^{ab}_{kl} +
               f^{l}_{ki}\widehat{W}^{ab}_{jl}
\ea\eqn{result1}
The fact that the breaking term $A_{i}(\f)$ vanishes, means that the 
Ward operators 
$\widehat{W}_{i} = \intg \widehat{W}^{a}_{i}\fud{}{\f^{a}}$
describe symmetries of the classical action $I(\f)$, by virtue of 
\equ{aexpr}
Moreover, from the expression \equ{xaexpr} for $X^{a}(C,\f)$ we 
derive the algebra compatible with the absorption of the ghost $\lambda$ 
through a redefinition of $C^{i}$~:
\eq
[\widehat{W}_{i},\widehat{W}_{j}] = f^{k}_{ij} \widehat{W}_{k}
+2\intg \widehat{W}_{ij}^{ab}\fud{I}{\f^{a}}\fud{}{\f^{b}}
\eqn{algebra1}
As it can be seen, \equ{algebra1} describes an on--shell algebra for 
the Ward operators $\widehat{W}_{i}$, whose structure constants are 
$f^{i}_{jk}$.

Following the same method, one finds that the ghost $\lambda$ can also be 
absorbed through a redefinition of the external fields as in 
\equ{ghat} and correspondingly the action $\widehat{\G}^{cl}$ has the 
form \equ{gammahat} with
\eq\ba{l}
W^{a}_{i} = \widehat{W}^{a}_{i} \es
W_{ij}^{ab} = \widehat{W}_{ij}^{ab} \es
X_{ij}^{a} =    \half \Lp f^{a}_{bi}  \widehat{W}^{b}_{j} -
                        f^{a}_{bj}  \widehat{W}^{b}_{i} \Rp  \es
A_{i} = 0 \es
X^{ab}_{ijk} = f^{a}_{ci}\widehat{W}^{bc}_{jk} 
\ea\eqn{result2}
Again, the $\widehat{W}_{i}$ are symmetries of $I(\f)$ and satisfy the 
algebra
\eq
[\widehat{W}_{i},\widehat{W}_{j}] = 
\intg \LP f^{a}_{bi} \widehat{W}^{b}_{j}\fud{}{\f^{a}}
+2\widehat{W}^{ab} \fud{I}{\f^{a}}\fud{}{\f^{b}} \RP
\eqn{algebra2}
which differs from the previous obtained one \equ{algebra1}, as 
different are the structures constants $f^{a}_{bi}$ as well.

Eliminating finally the ghost $\lambda$ from the theory by redefining both 
the ghosts $C^{i}$ and the external sources $\g_{a}$, by means of 
\equ{chat} and \equ{ghat}, one still finds that the $\widehat{W}_{i}$ 
must be symmetries of the action $I(\f)$, $i.e.$ $A_{i}(\f)=0$, and 
the algebra is
\eq
[\widehat{W}_{i},\widehat{W}_{j}] = f^{k}_{ij} \widehat{W}_{k}
+ \intg \LP 
f^{a}_{bi} \widehat{W}^{b}_{j}\fud{}{\f^{a}}
+2\widehat{W}^{ab} \fud{I}{\f^{a}}\fud{}{\f^{b}} \RP
\eqn{algebra3}
As one would expect, \equ{algebra3} is the direct product of the 
algebras~\equ{algebra1} and~\equ{algebra2}.

We conclude this Section with a remark concerning the 
cohomological structure of the model we are considering, with and 
without the constant ghost $\lambda$. In presence of $\lambda$, the 
cohomology of the linearized ST operator \equ{slavnovlin} is 
trivial~\cite{beppe}, 
although the algebraic constraints we find are not trivial. Otherwise, 
when $\lambda$ is not present because can be absorbed as it has been 
discussed, the cohomology depends on the detailed 
structure of the algebra one recovers. In other words, what is not 
trivial, is the cohomology of the ST operator constrained to its 
$\lambda$--independent sector.
That the presence of $\lambda$ 
somehow trivializes the cohomology of the theory can be easily seen, for 
instance, by using Dixon's filtration theorem~\cite{filtr}, which 
states that the cohomology of a nilpotent operator $s$ is isomorphic 
to a subspace of the cohomology of $s^{(0)}$, where $s^{(0)}$ is obtained 
by filtering $s$ with a filtration operator $\NN$, such that
\eq
s = \sum_{n\geq 0} s^{(n)}\ ,\ \ \ [\NN,s^{(n)}] = n s^{(n)}
\eqn{filterthm}
In our case, a convenient choice for $\NN$ is
\eq
\NN = \intg \f^{a}\fud{}{\f^{a}} + C^{i}\pad{}{C^{i}}
\eqn{filter}
according to which the filtered linearized ST operator 
\equ{slavnovlin} at the lowest order is
\eq
\BB_{\G^{cl}}^{(0)} = \pad{}{\lambda}
\eqn{b0}
which is clearly a nilpotent operator with empty cohomology. Hence, 
the cohomology of the whole operator $\BB_{\G^{cl}}$ is empty, too.

\section{Conclusions}\label{sez5}
We considered a theory characterized by an action $I(\f)$ and by some 
nonlinear Ward operators $\W$\ . We put ourselves in the most general 
case, with the $\W$ neither describing symmetries nor forming a 
closed algebra, being motivated by the fact that theories of physical 
relevance, like supersymmetric and topological models for instance, 
are of this type. 

At the tree level, we found the most general 1PI generating 
functional $\G^{cl}$ embedding the action $I(\f)$ which satisfies the 
Slavnov--Taylor identity \equ{slavnovid}  built from the $\W$'s and 
depending on the quantum fields $\phi_{a}(x)$, the sources 
$\gamma_{a}(x)$ and two types of global ghosts: the $C^{i}$ and 
$\lambda$, associated to $\W$ and to the breaking of 
the algebra, respectively. Finally, everything turns out to be 
expressed  in terms of the 
known quantities of the theory, {\it i.e.} the action $I(\f)$ and the Ward 
operators $\W$, and of the integrability condition~\equ{wexpr}.
Our formalism has been explicitly applied to the case of  
$N=4$ supersymmetric nonlinear sigma model, which shows an open 
algebra of the type described in this paper. In this context, the 
general integrability condition \equ{wexpr}, and hence the 
Slavnov--Taylor identity which originates it, 
turns out to have a remarkable geometrical meaning.

In addition, by means of equation~\equ{dexpr}, we determined the structure of 
the breaking of the 
algebra compatible with the Slavnov--Taylor identity, and thus with 
the quantization of the theory. 

As last step, we studied the possibility of eliminating the ghost 
$\lambda$, by redefining the ghosts $C^{i}$ and/or the sources 
$\g_{a}(x)$. We found that this is possible for various types of algebras, 
broken only by terms vanishing on--shell, and for Ward operators $\W$ 
which, differently from the general case analyzed in this paper,  
describe symmetries of the theory. 
\vspace{12mm}

\noindent   {\bf Acknowledgments}: It is a pleasure to thank 
B.Bandelloni for discussions and for his comments.


\end{document}